\documentclass[a4paper]{article}
\usepackage{epsfig,amsmath}

\newcommand{\tesla}{\textsc{Tesla}\ }
\newcommand{\gev}{{\rm \ GeV}}
\newcommand{\mev}{{\rm \ MeV}}
\newcommand{\re}{\Re e \,}

\newcommand{\Eslash}{{\not{\!\!E}}}
\newcommand{\MW}{M_{\rm W}}
\newcommand{\dMW}{\delta M_{\rm W}^2}
\newcommand{\MZ}{M_{\rm Z}}
\newcommand{\sw}{s_{\rm W}}
\newcommand{\cw}{c_{\rm W}}

\newcommand{\anc}{\rule{0mm}{0mm}}

\newcommand{\sll}{{\tilde{l}}}
\newcommand{\slR}{{\tilde{l}_{\rm R}}}
\newcommand{\slL}{{\tilde{l}_{\rm L}}}
\newcommand{\msl}{m_{\tilde{l}}}
\newcommand{\smu}{{\tilde{\mu}}}
\newcommand{\smuR}{{\tilde{\mu}_{\rm R}}}

\newcommand{\se}{{\tilde{e}}}
\newcommand{\seR}{{\tilde{e}_{\rm R}}}
\newcommand{\seL}{{\tilde{e}_{\rm L}}}
\newcommand{\mseR}{m_{\tilde{e}_{\rm R}}}
\newcommand{\mseL}{m_{\tilde{e}_{\rm L}}}
\newcommand{\msmuR}{m_{\tilde{\mu}_{\rm R}}}
\newcommand{\cha}{\tilde{\chi}}
\newcommand{\neu}{\tilde{\chi}^0}
\newcommand{\mcha}[1]{m_{\tilde{\chi}^\pm_{#1}}}
\newcommand{\mneu}[1]{m_{\tilde{\chi}^0_{#1}}}

\newcommand{\sF}{{\tilde{f}}}

\newcommand{\PL}{{\rm L}}
\newcommand{\PR}{{\rm R}}
\newcommand{\PS}{{\rm S}}

\newcommand{\seCL}{\Sigma^{\pm\PL}}
\newcommand{\seCR}{\Sigma^{\pm\PR}}
\newcommand{\seCS}{\Sigma^{\pm\PS}}

\newcommand{\seNL}{\Sigma^{0\PL}}

\newcommand{\seNS}{\Sigma^{0\PS}}

\begin{document}

\begin{flushright}
DESY--02--182\\   
FERMILAB--Conf--02/270--T\\[2em]
\end{flushright}

\begin{center}

{\Large \textbf{Precise predictions for slepton pair production}%
\footnote{Presented by A.F. at the 10th International Conference on
Supersymmetry and Unification of Fundamental Interactions (SUSY02), June 17--23,
2002, DESY Hamburg, Germany, to appear in the proceedings.}
}\\[2ex]
{\large Ayres~Freitas$^{a,b}$\footnote{afreitas@fnal.gov}
,
Andreas~v.~Manteuffel$^{b}$}\\[1ex]
{\it $^a$Fermi National Accelerator Laboratory, Batavia, IL 60510-500, USA\\
$^b$Deutsches Elektronen--Synchrotron DESY, D--22603 Hamburg, Germany}

\end{center}

\begin{abstract}
\noindent
At a future linear collider, the masses and couplings of scalar leptons can be
measured with high accuracy, thus requiring precise theoretical predictions for
the relevant processes.
In this work, after a discussion of the expected experimental precision, the
complete one-loop corrections to smuon and selectron pair production in the
MSSM are presented and the effect of different contributions in the result is
analyzed.
\end{abstract}

\section{Introduction}

If supersymmetric particles are detected in the future, their properties can be
studied with high accuracy at a high-energy linear collider \cite{LC}.
Accordingly, precise theoretical predictions for the anticipated processes are
important for the determination of the couplings and the underlying
supersymmetry-breaking parameters.
In this report, the production of scalar leptons
near threshold and in the continuum is analyzed, focusing on the production of
scalar electrons (selectrons) $\se$ and scalar muons (smuons) $\smu$ in $e^+e^-$
annihilation. The possibility to produce selectrons in the $e^-e^-$ mode is also
discussed.

The status of theoretical predictions for the excitation curves near threshold,
being relevant for the measurement of the slepton masses, is shortly summarized.
The measurement of the cross-sections in the continuum, on the other hand, can
be used to precisely determine the couplings of the sleptons. For this purpose,
the complete next-to-leading order radiative corrections to the production of
selectrons and smuons in the framework of the Minimal
Supersymmetric Standard Model (MSSM) are presented. Numerical results are given
for the ${\cal O}(\alpha)$ corrections to the processes $e^+e^- \to
\smuR^+\smuR^-$ and $e^\pm e^- \to \seR^\pm\seR^-$.

\section{Precision measurements near threshold}

At a linear collider with high luminosity, the masses of sleptons can be
determined with high accuracy by measuring the shape of the production
cross-section near threshold. Previous analyses have shown that the expected
experimental precision for the mass measurement is of the order ${\cal O}(100
\mev)$ \cite{LC}. It is therefore necessary to incorporate effects beyond leading
order in the theoretical predictions in order to match the experimental
accuracy. Near threshold, important corrections to the cross-sections arise from
the non-zero slepton width and the Coulomb corrections \cite{thr1}. 
The slepton width $\Gamma_{\sll}$ is expected to be small compared to the
slepton mass $\msl$.
It can be incorporated by introducing a complex mass for the
intermediate off-shell sleptons, $\msl^2 \to \msl^2 - i \msl \Gamma_{\sll}$. 
The Coulomb
rescattering correction is one of the most important radiative corrections near
threshold arising from photon exchange between the slowly moving sleptons. 
For the production of off-shell sleptons with orbital angular
momentum $l$ its leading contribution is given by
\vspace{-1pc}
\begin{equation}
\sigma_{\rm coul} =  \sigma_{\rm born} \,
\frac{\alpha\pi}{2 \beta_p} Q^2_{\sF} 
\biggl [ 1\! -\! \frac{2}{\pi} \arctan
\frac{|{\beta_M}|^2 - \beta_p^2}{2 \beta_p \;
        \Im \! m \, \beta_M} \biggr ] 
 \,\Re e
  \biggl [\frac{\beta_p^2 + \beta_M^2}{2 \beta_p^2}
  \biggr ]^l
\end{equation}
with
$
\beta_p = \displaystyle\frac{1}{s}\sqrt{(s-p_+^2-p_-^2)^2-4
p_+^2 p_-^2}, 
$ and $
\beta_M = \displaystyle\frac{1}{s}\sqrt{(s-M_+^2-M_-^2)^2-4
M_+^2 M_-^2}.
$ 
Here $Q_{\sF}$, $p_\pm$ and $M^2_\pm = m_\pm^2 - i \, m_\pm \Gamma_{\!\pm}$ are
the charge, the momenta and complex pole masses of the
slepton and anti-slepton.

The slepton signal is characterized by their decays into neutralinos
and charginos. Here the decay channels $\slR^- \to l^- \neu_1$ and $\slL^- \to l^-
\neu_2 \to l^-\tau\tau \neu_1$ have been considered, with the lightest
neutralino $\neu_1$ escaping undetected.
Important backgrounds arise both from both Standard Model and supersymmetric
sources.
After applying appropriate cuts to reduce the background, the slepton masses and widths
can be extracted in a model-independent way from the measurement of the
threshold cross-section.
In Tab.~\ref{tab:thrmass} results are given 
from a fit to cross-section measurements at five equidistant center-of-mass
energies near threshold.
%
%
%
%
The cross-sections have been simulated including the aforementioned
theoretical corrections as well as initial-state radiation in the leading-log approximation and
beamstrahlung effects.
\begin{table}[bt]
\caption{Expected precision for the determination of selectron and smuon
masses and widths
from measuring the threshold cross-sections at five equidistant points.
The values are for the SPS1 scenario~\cite{sps}.}
\label{tab:thrmass}
\vspace{1ex}
\setlength{\tabcolsep}{5mm}
\renewcommand{\arraystretch}{1.3}
\begin{tabular*}{\columnwidth}{@{}p{5cm}@{}cll@{}}
\hline
Process & Integr. Lumin. [fb$^{-1}$] & \multicolumn{1}{c}{Mass [GeV]} & Width [MeV] \\
\hline
$e^+e^- \!\!\to (\seR^+ \seR^-)
\to e^+ e^- + \Eslash$ & $5\!\times\! 10$ & $\mseR = 143.0^{+0.21}_{-0.19}$ &
  $\Gamma_{\!\seR} = 150^{+300}_{-250}$ \\
$e^-e^- \!\!\to (\seR^- \seR^-)
\to e^- e^- + \Eslash$ & $5\!\times\! 1$ & $\mseR = 142.95^{+0.048}_{-0.053}$ &
  $\Gamma_{\!\seR} = 200^{+50}_{-40}$ \\
\hline
$e^+e^- \!\!\to (\seR^\pm \seL^\pm)\to e^+ e^- \tau\tau +\Eslash$ &
  $5\!\times\! 10$ & $\mseL = 202.2^{+0.37}_{-0.33}$ &
  $\Gamma_{\!\seL} = 240^{+40}_{-40}$ \\
$e^-e^- \!\!\to (\seL^- \seL^-)\to e^- e^- 4\tau +\Eslash$ &
  $5\!\times\! 1$ & $\mseL = 202.2^{+0.62}_{-0.44}$ &
  $\Gamma_{\!\seL} = 240^{+500}_{-400}$ \\
\hline
$e^+e^- \!\!\to (\smuR^+ \smuR^-)
\to \mu^+ \mu^- + \Eslash$ & $5\!\times\! 10$ & $\msmuR = 143.0^{+0.42}_{-0.38}$ &
  $\Gamma_{\!\smuR} = 350^{+400}_{-400}$ \\
\hline
\end{tabular*}
\end{table}
The findings of this study are consistent with the numbers quoted in
\cite{martynco}, which have been obtained without taking into account
background contributions and partly with higher integrated luminosity.

\section{Analysis of slepton couplings}

\begin{figure}[tb]
\anc\\[1ex]
\anc (a) \hspace{8cm} (b) \\[-1ex]
\anc \hspace{2em}
\begin{minipage}[b]{6cm}
\psfig{figure=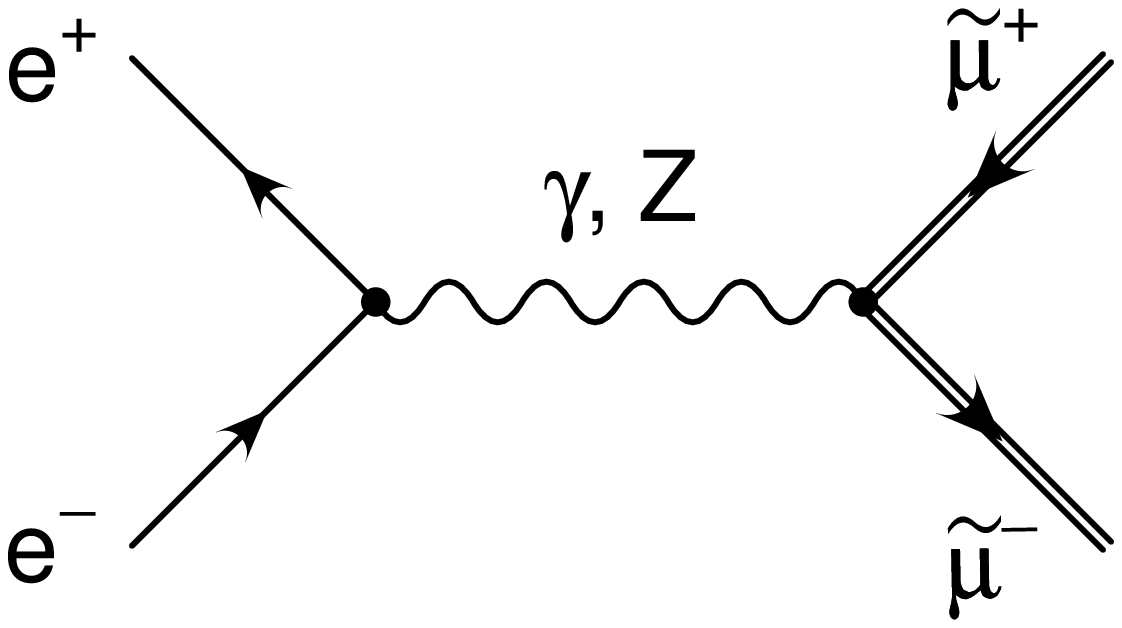, width=5cm}\\
\vspace{-2.28cm}\\ \anc\hspace{3.5cm}
{\LARGE
$\bullet$}\hspace{3mm}$\longleftarrow\bar{g}_1,\bar{g}_2$
\end{minipage}
\hspace{2cm}
\begin{minipage}[b]{6cm}
\psfig{figure=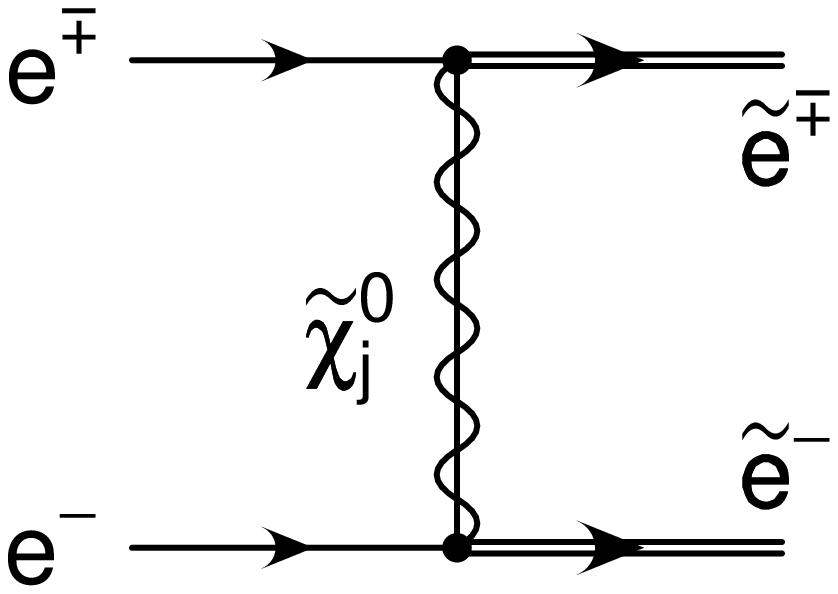, width=4.3cm}\\
\vspace{-3cm}\\ \anc\hspace{2.2cm}\parbox{2cm}{
{\LARGE$\bullet$\\}
\anc\hspace{5mm}$\nwarrow$\\[.5mm]
\anc\hspace{8mm}$\hat{g}_1, \hat{g}_2$\\[1mm]
\anc\hspace{5mm}$\swarrow$\\[1mm]
{\LARGE$\bullet$}
}
\end{minipage}
\caption{Slepton gauge and Yukawa couplings in tree-level contributions to smuon
(a) and selectron (b) pair production. $\bar{g}_1$ and $\bar{g}_2$ denote the
slepton U(1)$_{\rm Y}$ and SU(2)$_{\rm L}$ gauge coupling, while $\hat{g}_1$ and
$\hat{g}_2$ indicate the corresponding Yukawa couplings.}
\label{fig:cdiag}
\end{figure}

In contrast to the masses of the superpartners, their couplings are not directly
modified by soft-breaking terms. As a consequence, supersymmetry relates the
Standard Model gauge coupling between a vector boson $V$ and a
fermion $f$, $g(Vff)$ to the gauge coupling
between the vector boson $V$ and the scalar fermion $\tilde{f}$,
$\bar{g}(V\tilde{f}\tilde{f})$, as well as to
the Yukawa coupling between the
gaugino partner $\tilde{V}$ of the vector boson, the fermion $f$ and the
sfermion $\tilde{f}$, $\hat{g}(\tilde{V}f\tilde{f})$. Within softly broken
supersymmetric theories all three  kinds of
couplings are required to be identical,
$
g = \bar{g} = \hat{g}.
$
The gauge and Yukawa couplings of scalar fermions can be precisely determined by
measuring their production cross-sections at a linear collider. For the case
of the SU(3) QCD sector in the MSSM this has been investigated in
Ref.~\cite{susyqcd}.
In the electroweak sector, which comprises the hypercharge U(1)$_{\rm Y}$ coupling
$g_1$ and the SU(2)$_{\rm L}$ coupling $g_2$, the coupling relation can be accurately
tested by measuring the pair-production cross-sections of scalar leptons.
The pair production for smuons, which proceeds via s-channel photon and
Z-boson  exchange, is particularly suited for the extraction of the slepton
gauge couplings $\bar{g}_{1,2}$, see Fig.~\ref{fig:cdiag}~(a).
On the other hand, the Yukawa couplings $\hat{g}_{1,2}$
can be probed best in the production of selectrons, as a result of the t-channel
neutralino exchange, see Fig.~\ref{fig:cdiag}~(b). As mentioned above, selectron
production can also be studied in $e^-e^-$ collisions.

In order to extract the Yukawa couplings from the measurement of the selectron
cross-sections, it is necessary to use information about the neutralino system.
Here it is assumed that the neutralino sector has the form of the MSSM,
essentially depending on the gaugino parameters $M_1$, $M_2$ and the Higgs
parameter $\mu$. The dependence on $\tan\beta = v_{\rm
u}/v_{\rm d}$, the ratio of the vacuum expectation values of the two Higgs
doublets, is relatively mild and can be neglected if the value of $\tan\beta$
can be extracted with moderate accuracy from some other measurement like Higgs
decay branching ratios. Thus the three parameters $M_1$, $M_2$ and $\mu$ can be
determined from the measurement of three chargino or neutralino masses. Here
the two chargino masses and the lightest neutralino mass have been used,
assuming the following---rather conservative---experimental errors:
$\delta \mcha{1} = 100 \gev$, $\delta \mcha{2} = 400 \gev$, $\delta \mneu{1} =
100 \gev$.
The total slepton cross-sections and signal-to-background ratios can be enhanced
by suitable beam polarization. It is assumed that the polarization degree can
be controlled up to 1\%. The backgrounds are further reduced by applying
appropriate cuts \cite{thr1, mythesis}. As before, the same decay modes and
final state signatures listed in the first column of Tab.~\ref{tab:thrmass} have
been used.

\begin{figure}[t]
\centering{
\begin{tabular}{l@{\hspace{1cm}}l}
 \underline{(a) $e^+e^-$, $\sqrt{s} = 500 \gev$, $L = 500$ fb$^{-1}$} &
 \underline{(b) $e^-e^-$, $\sqrt{s} = 500 \gev$, $L = 50$ fb$^{-1}$}
 \\[2ex]
\hspace{-1cm}
\epsfig{file=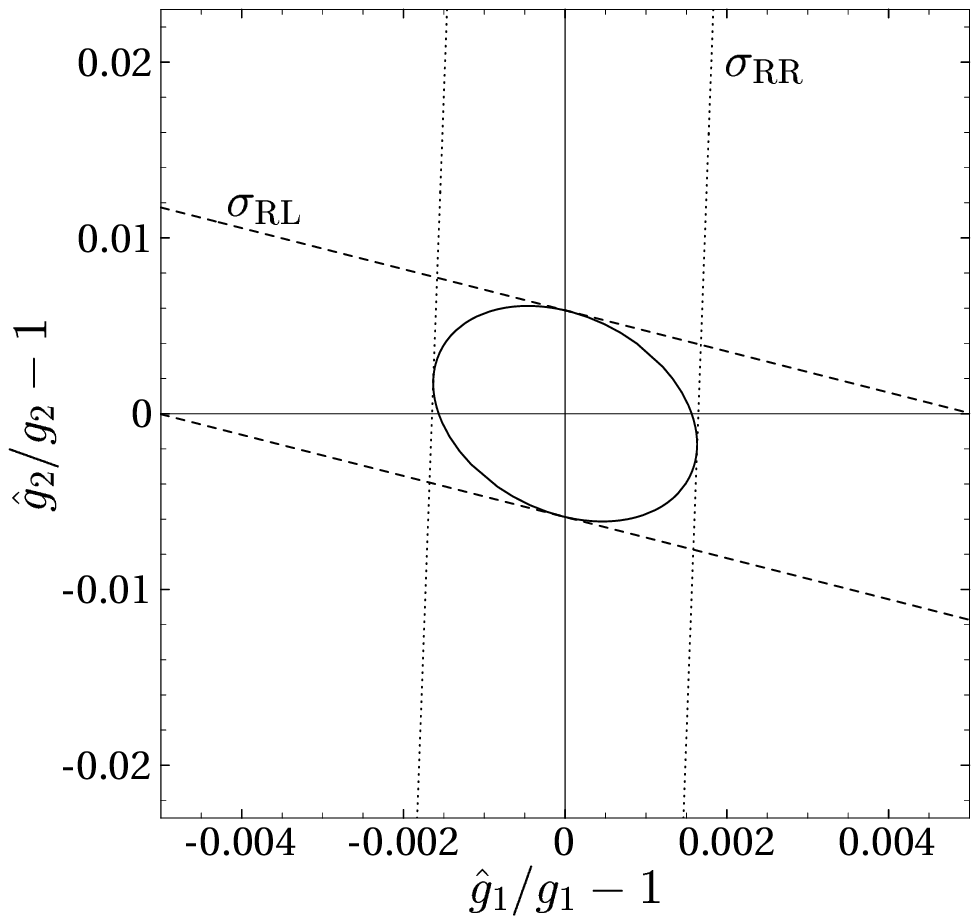,width=8.2cm}&
\hspace{-1cm}
\epsfig{file=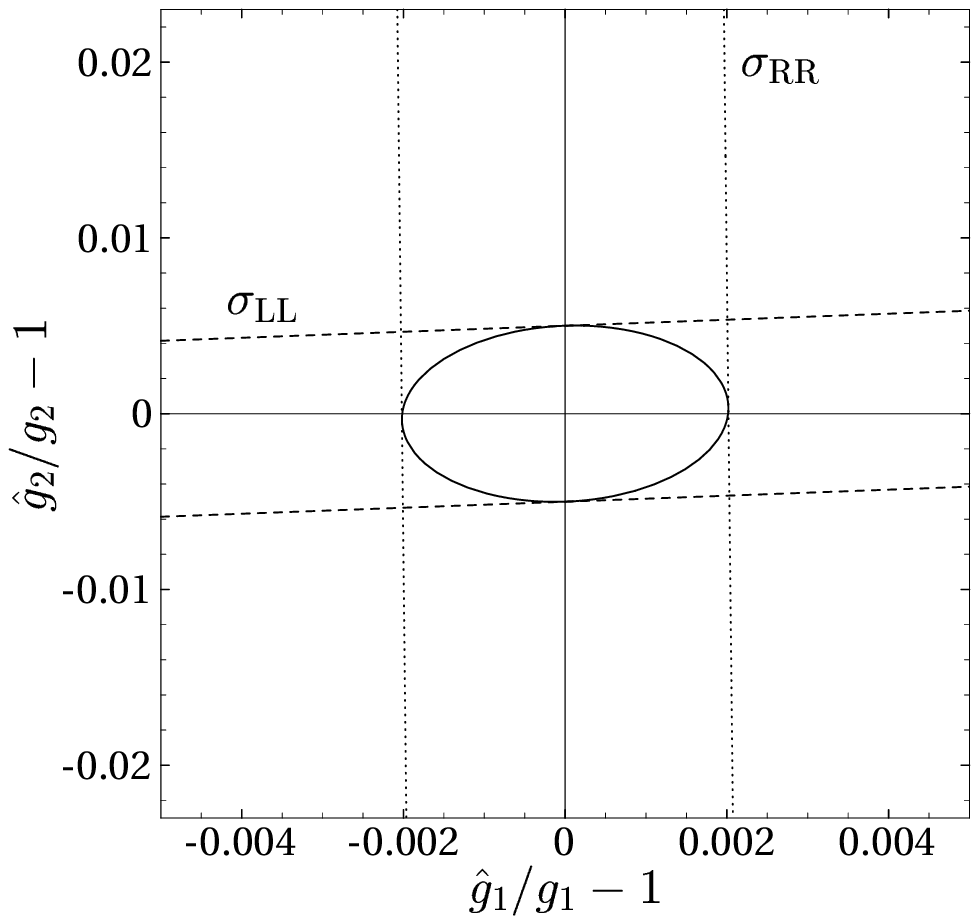,width=8.2cm}
\end{tabular}				
}
\vspace{-1em}
\caption{1$\sigma$ bounds on the determination of the supersymmetric Yukawa
couplings $\hat{g}_1$ and $\hat{g}_2$ from selectron cross-section measurements.
The two plots compare the information obtained from
the cross-sections $\sigma_{\rm RR} =
\sigma[e^+ \, e^- \to \seR^+ \, \seR^-]$ and $\sigma_{\rm RL} =
\sigma[e^+ \, e^- \to \seR^\pm \, \seL^\mp]$ in the $e^+e^-$ mode (a)
as well as
$\sigma_{\rm RR} =
\sigma[e^- \, e^- \to \seR^- \, \seR^-]$ and $\sigma_{\rm LL} =
\sigma[e^- \, e^- \to \seL^- \, \seL^-]$ in the $e^-e^-$ mode, respectively.
The values are for the SPS1 scenario \cite{sps}.
}
\label{fig:ggp}
\end{figure}

The resulting constraints on the Yukawa couplings $\hat{g}_1$ and
$\hat{g}_2$ from selectron cross-section measurements are depicted in
Fig.~\ref{fig:ggp}.
From the figure the following resulting accuracies are obtained:
\begin{eqnarray}
e^+e^-: && \delta \hat{g}_1 / \hat{g}_1 \approx 0.18\%, \qquad
		\delta \hat{g}_2 / \hat{g}_2 \approx 0.8\%, \\
e^-e^-: && \delta \hat{g}_1 / \hat{g}_1 \approx 0.23\%, \qquad
		\delta \hat{g}_2 / \hat{g}_2 \approx 0.8\%.
\end{eqnarray}
On the other hand, from the measurement of the R-smuon production cross-section,
the smuon gauge coupling can be extracted with a total error of 1\%.

Therefore it is clearly necessary to include radiative corrections in the
theoretical predictions for the slepton cross-sections in order to
match the experimental accuracy. 
As a
first step the produced sleptons may be considered on-shell, since in the
continuum far above threshold the effect of the non-zero slepton width
on the total cross-section
is relatively small, of the order $\Gamma_{\!\sll}/\msl$.
Thus,
the production and decay of the sleptons
can be treated separately. 
While the MSSM one-loop corrections to the decay of
sleptons into neutralinos and charginos, $\sll^\pm \to l^\pm \, \neu_i$ and
$\sll^\pm \to ^{^{_{(-)}}} \hspace{-3ex} \nu_l \, \cha^\pm_i$,
have been computed in Refs.~\cite{Guasch:01, Guasch:02},
the following sections will discuss the complete next-to-leading order
corrections in the MSSM to
the production of scalar leptons.

\section{Radiative corrections and renormalization}

The computation of the ${\cal O}(\alpha)$ corrections to scalar lepton
production requires the inclusion of all sectors of the electroweak MSSM
in the loop contributions. In order to reduce
the number of parameters the following simplifications have been made:

All soft-breaking parameters of the MSSM are assumed to be real, i.e. only the
case of CP conservation is considered. The masses of the fermions of the first
two generations are neglected, and accordingly no mixing between the left- and
right-chiral components of the first and second generation sfermions occurs. On
the other hand, due to the large Yukawa couplings, the fermion masses and
sfermion mixings in the third generation are fully taken into account. The CKM
matrix is assumed to be diagonal and no flavour mixing between the sfermions is
considered.

In this work the on-shell renormalization scheme has been used,
which relates the mass parameters to the pole position of the propagators and
the electric charge to the electron coupling in the Thomson limit.
The gauge sector of the MSSM is renormalized similar to the Standard Model gauge
sector. The relevant expressions can be found e.g. in Ref.~\cite{Denner:93}. For
the production of scalar leptons of the first two generations, mixing between
the sleptons can be neglected, as mentioned above. Accordingly, the L- and
R-sleptons can be renormalized independently.

A large number of MSSM couplings and masses depend on $\tan\beta = v_{\rm
u}/v_{\rm d}$. However, the vacuum expectation values and $\tan\beta$ are not
physical quantities, so that it is difficult to relate $\tan\beta$ to an
observable \cite{stoecki:02}. For technical reasons it is instead advantageous
to renormalize $\tan\beta$ in the $\overline{\rm DR}$ scheme, which amounts to
cancelling only the divergent part in dimensional reduction with the
counterterm. This definition of $\tan\beta$ has been used here.

The mass spectrum of the two chargino and four neutralino states depends only
on three independent parameters, the superpotential parameter $\mu$ and the
gaugino parameters $M_1$ and $M_2$. In our renormalization procedure,
counterterms for these parameters are introduced, which enter in the
renormalization of the chargino and neutralino mass matrices. The counterterms
$\delta M_1$, $\delta M_2$, $\delta\mu$ are fixed by imposing on-shell
conditions for three of the six mass eigenvalues \cite{Pierce, Fritzsche:02}%
\footnote{An alternative renormalization procedure has been given in
Ref.~\cite{Eberl:01} which is physically equivalent at the one-loop level, but
differs in the definition of the renormalized parameters $M_1, M_2, \mu$.},
for which here the two chargino masses and the lightest neutralino mass have
been chosen:
\begin{align}
\delta M_2 &=
  \frac{1}{\mu^2-M_2^2} \Bigl [ (\mcha{2} \mu - \mcha{1} M_2)
  \; \delta \mcha{1} + (\mcha{1} \mu - \mcha{2} M_2) \; \delta \mcha{2}
+ M_2 \; \dMW + \mu \; \delta \bigl ( \MW^2 \sin 2 \beta \bigr )  \Bigr],
\nonumber \\
\delta \mu &= 
  \frac{1}{M_2^2-\mu^2} \Bigl [ (\mcha{2} M_2 - \mcha{1} \mu)
  \; \delta \mcha{1} + (\mcha{1} M_2 - \mcha{2} \mu) \; \delta \mcha{2}
+ \mu \; \dMW + M_2 \; \delta \bigl ( \MW^2 \sin 2 \beta \bigr )  \Bigr],
\nonumber \\
&\text{with } \quad \delta \mcha{k} = \frac{1}{2} \,
  \re \! \bigl\{ \mcha{k} \seCL_{k}(\mcha{k}^2) +
  \mcha{k} \seCR_{k}(\mcha{k}^2) + 2\, \seCS_{k}(\mcha{k}^2) \bigr\},
  \nonumber \\
\delta M_1 &= \begin{aligned}[t]
  \frac{1}{N_{11}^2} \Bigl [ &\re \! \bigl\{
  \mneu{1} \seNL_{1}(\mneu{1}^2) + \seNS_{1}(\mneu{1}^2) \bigr \}
  - N_{12}^2 \, \delta M_2 + 2 N_{13} N_{14} \, \delta\mu \\
+ \; &2 N_{11} \bigl [ N_{13} \; \delta \bigl (\MZ \sw \cos\beta \bigr )
                - N_{14} \; \delta \bigl (\MZ \sw \sin\beta \bigr ) \bigr ] \\
+ \; &2 N_{12} \bigl [ N_{13} \; \delta \bigl (\MZ \cw \cos\beta \bigr )
                - N_{14} \; \delta \bigl (\MZ \cw \sin\beta \bigr ) \bigr ]
\Bigr ]. \end{aligned} \label{eq:M1-ct}
\end{align}
Here $\Sigma_k^{\pm \rm L,R,S}$ and $\Sigma_k^{0 \rm L,R,S}$ are the left-,
right-handed and scalar components of the unrenormalized self-energy of the
$k$-th chargino and neutralino, respectively. $N$ denotes the neutralino mixing
matrix, which is taken at the tree-level since the above equations are only
required at the one-loop level.
For more details on the renormalization procedure, see
Refs.~\cite{mythesis,Guasch:02}.

The UV-divergent loop integrals have been regularized
using dimensional reduction \cite{dred}, which preserves gauge invariance
and supersymmetry at least up to the one-loop level. On the other hand, dimensional
regularization, being widely used for Standard Model calculations, is known to
violate supersymmetry, which therefore in general needs to be restored by extra
counterterms. However, since dimensional regularization preserves gauge
invariance, no symmetry-restoring counterterms are required for the
renormalization of gauge couplings and masses. In fact, since the
production of smuons only involves gauge couplings at tree-level, the one-loop
corrections can directly be computed both with dimensional regularization and
dimensional reduction. 
It has been explicitly checked that the results for smuon
production agree for both methods. In the loop corrections to selectron
production, on the other hand, there is a finite difference between the
two regularization schemes. Here dimensional reduction has been used.

The computation of the loop contributions was performed using the computer
algebra tools \textsl{FeynArts 3.0} \cite{feynarts} and \textsl{FeynCalc 2.2}
\cite{feyncalc}. Throughout the
calculation, a general $R_\xi$ gauge was used. By reducing the results to
generic scalar one-loop integrals, the gauge-parameter independence was
explicitly verified.

In order to obtain IR-finite and physically meaningful results, the virtual
corrections to both processes have been supplemented with the real photon
bremsstrahlung contributions. For the numerical analyses in the next section,
the real photon emission has been treated fully inclusive.

\section{Discussion of one-loop results for $\smu$ and $\se$ pair production}

In the following numerical results are presented in the SPS1 scenario
\cite{sps}. A large part of the next-to-leading order corrections stems from
universal QED contributions to the electromagnetic coupling and initial-state
photon radiation. The dominant contributions to the electromagnetic coupling
from light-fermion loops (i.e. all fermions but the top-quark)
can easily be incorporated by a shift
$\Delta\alpha$ of the fine structure constant $\alpha$. The radiation of soft
and collinear photons from the incoming $e^\pm$ leads to large logarithmic
corrections $\propto \log(Q/m_{\rm e})$ with $Q = {\cal O}(\sqrt{s})$. They can
be taken into account by convoluting the Born cross-section with a radiator
function. These universal terms are therefore dropped from the next-to-leading
order result. The effect of the non-universal residual corrections is then given
by
\begin{equation}
\Delta_\alpha = (\sigma_{\rm NLO} - \sigma_{\rm Born}) / \sigma_{\rm Born},
\end{equation}
where $\sigma_{\rm Born}$ is the Born cross-section including the universal QED
effects.

\begin{figure}[t]
\begin{minipage}{11cm}
\epsfig{figure=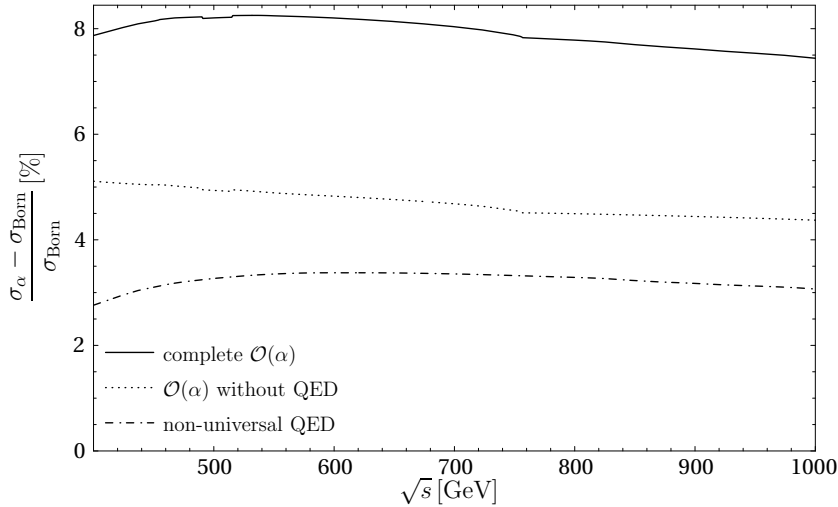, width=11cm}
\end{minipage}%
\hspace{5mm}%
\begin{minipage}{4.5cm}
\caption{Electroweak corrections to the cross-section for $e^+ e^- \to \smuR^+
\smuR^-$, relative to the Born cross-section.
Separately shown are 
full non-universal ${\cal O}(\alpha)$ contributions,
the genuine weak (non-QED) corrections and
the QED corrections, including soft and hard real
bremsstrahlung contributions but no universal ISR terms.
The input parameters are taken from SPS1
scenario~\cite{sps} with \mbox{$\msmuR = 143 \gev$}.}
\label{fig:smuon}
\end{minipage}
\end{figure}
\begin{figure}[t]
\begin{minipage}[b]{11cm}
(a)\\[-1em]
\epsfig{figure=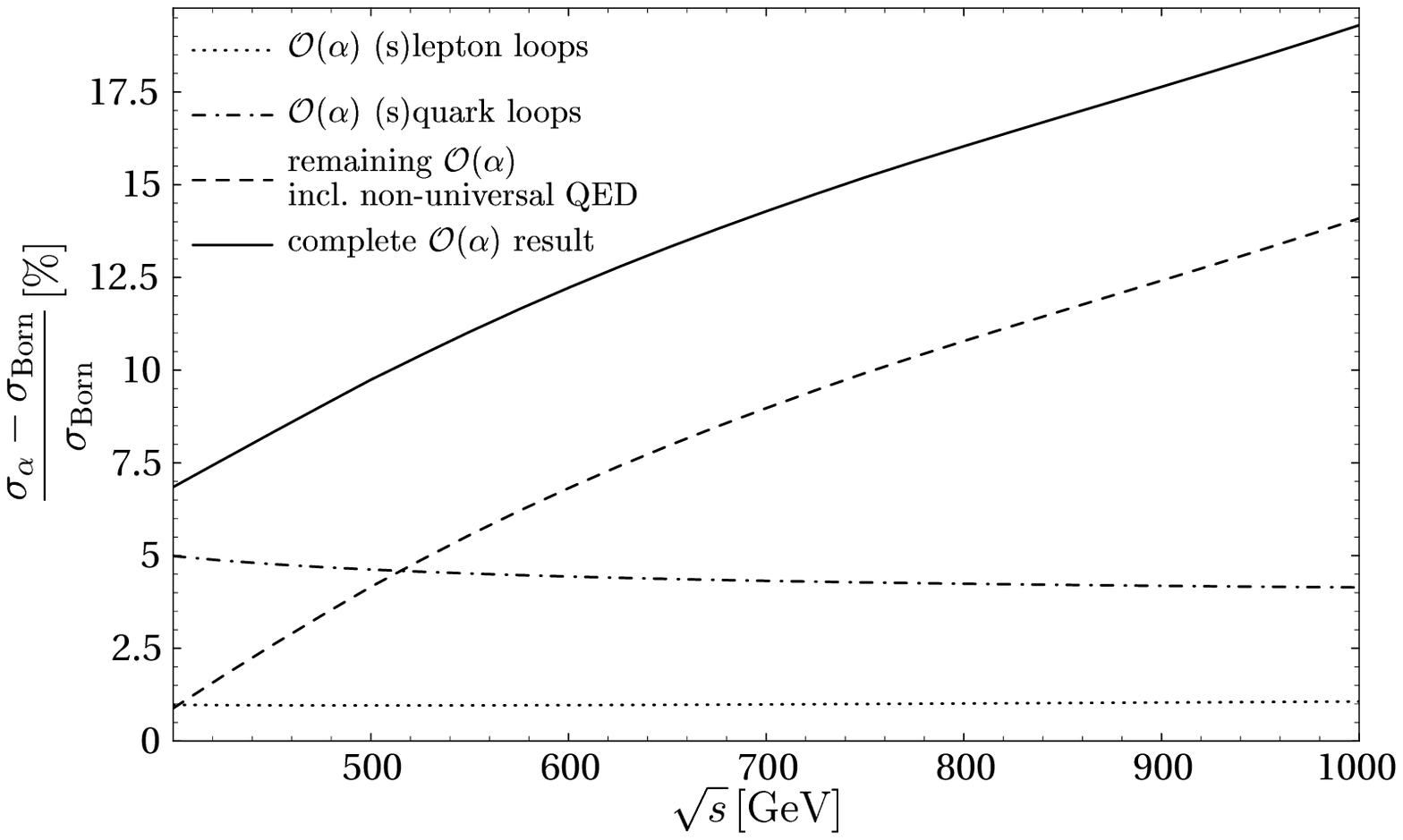, width=11cm, viewport=16 0 472 276}\\[1ex]
(b)\\[-1em]
\epsfig{figure=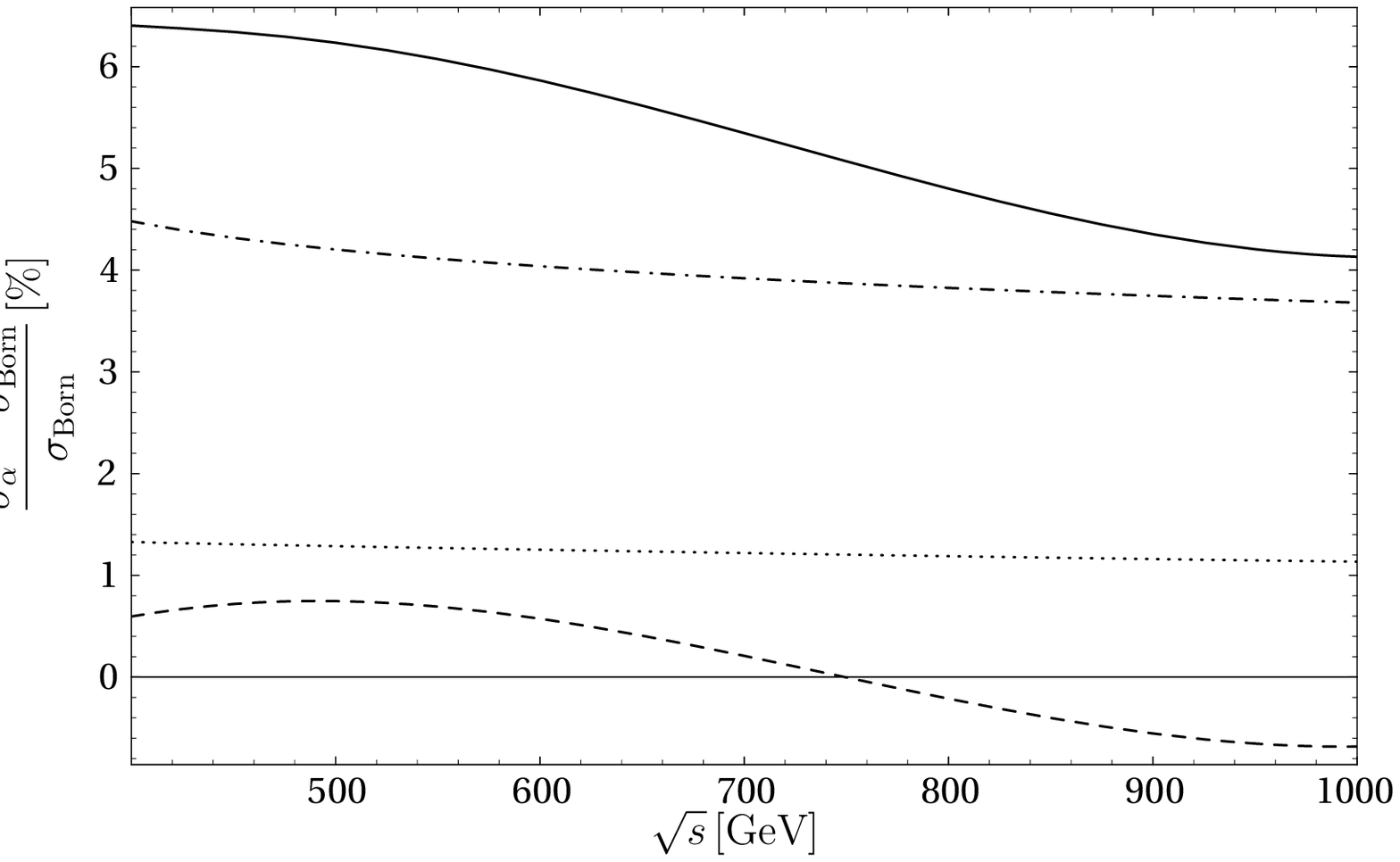, width=11cm}
\end{minipage}%
\hspace{5mm}%
\begin{minipage}[b]{4.5cm}
\caption{Electroweak corrections to the cross-section for $e^+ e^- \to \seR^+
\seR^-$ (a) and $e^- e^- \to \seR^-
\seR^-$ (b), relative to the Born cross-sections.
Besides the full non-universal 
${\cal O}(\alpha)$ results, contributions from different subsets of diagrams are shown.
The input parameters are taken from SPS1 scenario \cite{sps} with \mbox{$\mseR = 143
\gev$}.}
\label{fig:sel}\anc
\end{minipage}
\end{figure}

The effect of the remaining non-universal ${\cal O}(\alpha)$ corrections is
shown in Fig.~\ref{fig:smuon} for the production of R-smuons in $e^+e^-$
annihilation and in
Fig.~\ref{fig:sel}~(a) and (b) for the production of R-selectrons in $e^+e^-$ and
$e^-e^-$ collisions, respectively. The effect of the next-to-leading order
contributions amounts to 5--10\%.

An interesting feature of the loop corrections to selectron production is, in
contrast to smuon production, the non-decoupling behaviour of supersymmetric
particles in the loops. This is related to the fact that the tree-level
amplitude for smuon production only involves gauge couplings, see
Fig.~\ref{fig:cdiag}~(a), whereas the t-channel diagrams to selectron
production involve the electron-selectron-neutralino Yukawa coupling, cf.
Fig.~\ref{fig:cdiag}~(b). 
The origin of this effect can be illustrated by the renormalization group (RG)
running of the gauge and Yukawa coupling in the $\overline{\rm MS}$ or
$\overline{\rm DR}$ scheme, where the fundamental couplings are
modified in the low energy region by the non-decoupling contributions
from the high scale~\cite{superoblique}.
In the following, the effect arising
from quark/squark loops will be considered as an example.

Above the squark mass scale ($Q > m_{\rm \tilde{q}}$), supersymmetry is unbroken
so that the gauge coupling $g(Q)$ and the Yukawa coupling $\hat{g}(Q)$ are
equal. At $Q = m_{\rm \tilde{q}}$ the squarks decouple from the RG running of
the couplings. While $g(Q)$ still runs for $Q < m_{\rm \tilde{q}}$
because of quark loop contributions, there is no running of $\hat{g}(Q)$ from
quark/squark loops below $Q = m_{\rm \tilde{q}}$. When comparing the two
couplings at the weak scale, e.g. $Q = \MW$, they therefore differ by a
logarithmic term 
\begin{equation}
\hat{g}(Q)/g(Q) -1 \propto \log(m_{\rm \tilde{q}}/\MW). \label{eq:suob}
\end{equation}
It is obvious that this contribution does not vanish in the limit of large
squark masses.

\begin{figure}[t]
\begin{tabular}{@{}c@{$\,$}c@{}} 
\multicolumn{1}{@{}l}{\small $\Delta_\alpha$ [\%]} &
\multicolumn{1}{@{$\,$}l}{\small $\Delta_\alpha$ [\%]} \\[1.5ex]
 (a) & (b) \\[-1.8em]
\epsfig{file=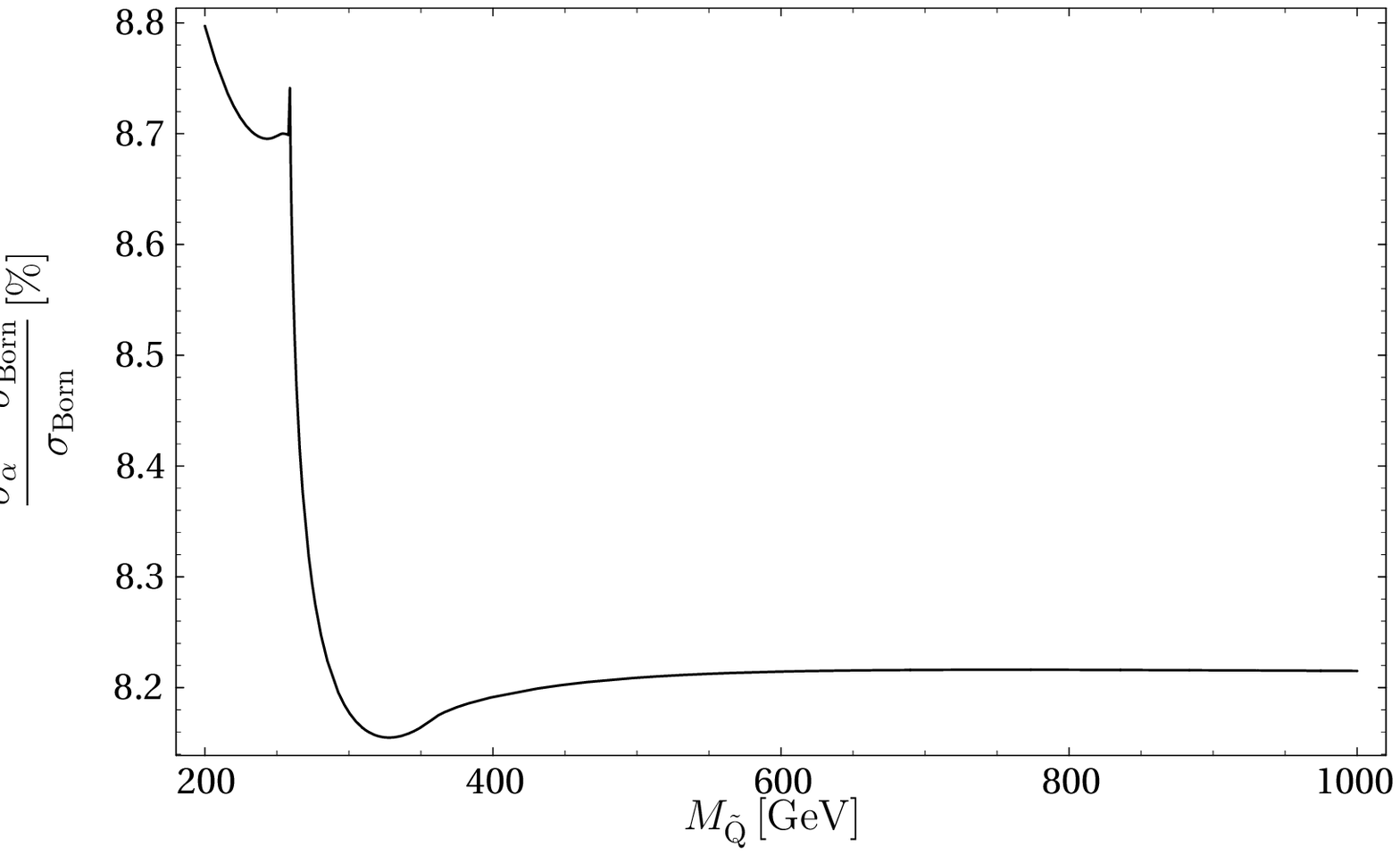,width=8cm, viewport=38 -3 472 282, clip=true} &
\epsfig{file=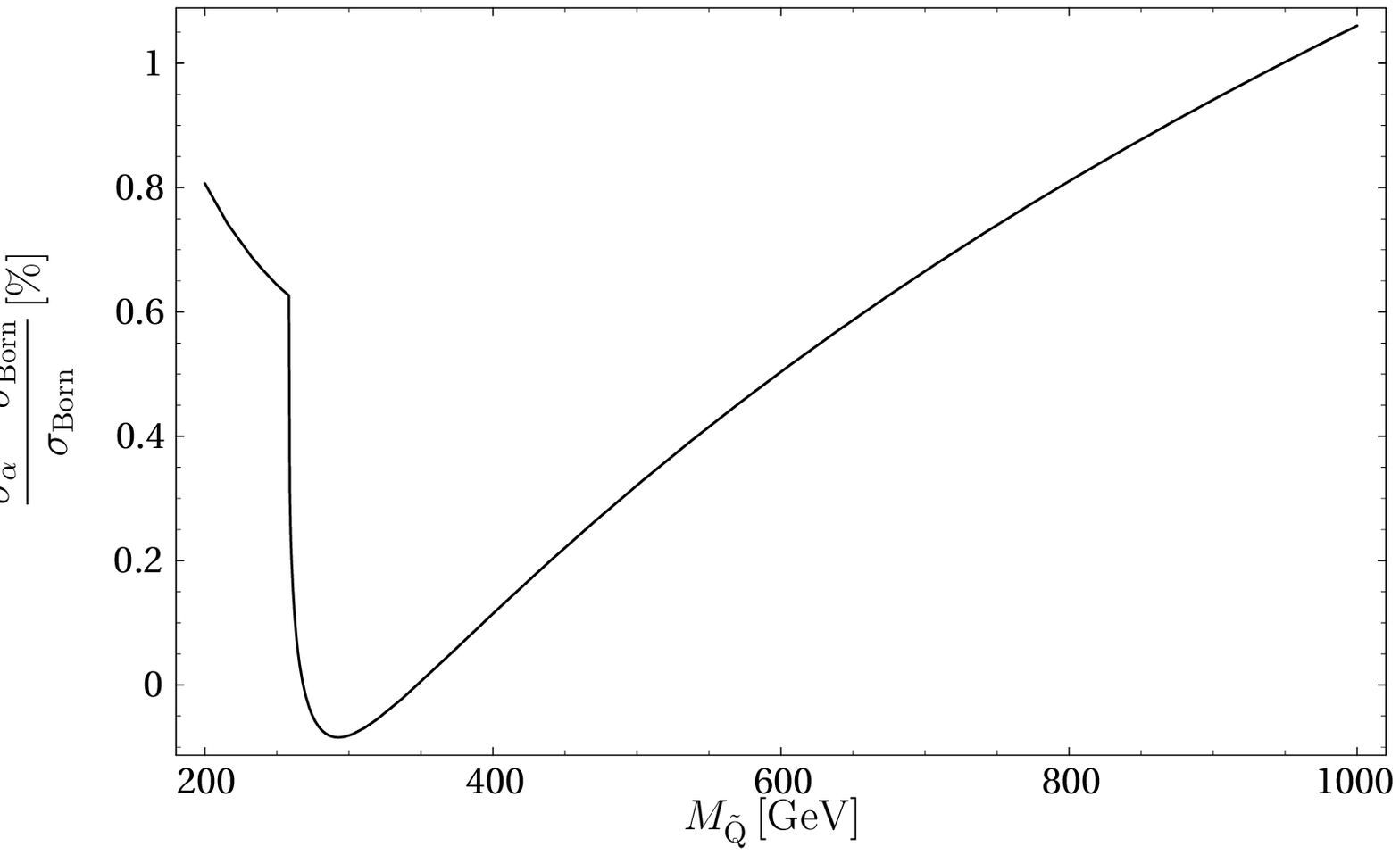,width=8cm, viewport=38 -3 472 282, clip=true}
\end{tabular}
\vspace{-1em}
\caption{Dependence of the relative corrections $\Delta_\alpha = (\sigma_{\rm
NLO} - \sigma_{\rm Born}) / \sigma_{\rm Born}$ [in \%] on the soft-breaking
squark mass parameter $M_{\rm \tilde{Q}}$ for $\smuR^+\smuR^-$ production (a)
and $\seR^-\seR^-$ production (b). The values of the other parameters are taken
from the SPS1 scenario \cite{sps} and $\sqrt{s} = 500 \gev$.
}
\label{fig:decoup}
\end{figure}

As mentioned before, the computation of the loop corrections to slepton
production in this work was performed in the on-shell scheme. The couplings in
this scheme are manifestly scale-invariant and obey the fundamental relation 
$g = \hat{g}$ at any scale in higher orders.
In the low-energy region, loop corrections
to the gauge bosons in gauge coupling vertices and to the
gauginos in gaugino-lepton-slepton vertices
lead to non-decoupling logarithmic contributions.
By comparing the resulting effective vertices
$g_{\rm eff}$ and $\hat{g}_{\rm eff}$ including one-loop corrections, one finds a corresponding relation to
eq.~\eqref{eq:suob},
\begin{equation}
\hat{g}_{\rm eff}/g_{\rm eff} -1 \propto \log(m_{\rm \tilde{q}}/\MW).
\end{equation}
The effect of the squark loop contributions can also be seen in
Fig.~\ref{fig:decoup}.
As evident from the figure, the $\smuR^+\smuR^-$ cross-section does not depend on the
common squark soft-breaking parameter $M_{\rm \tilde{Q}} = m_{\rm \tilde{q}_L} =
m_{\rm \tilde{u}_R} = m_{\rm \tilde{d}_R}$ for large values of
$M_{\rm \tilde{Q}}$. On the other hand, for increasing values of $M_{\rm
\tilde{Q}}$, the size of the radiative corrections to $\seR^-\seR^-$ production
grows logarithmically. For very large values of $M_{\rm
\tilde{Q}} \sim 100$ TeV, the effect of the squark loops can amount to a few
percent.

\vspace{2ex}\noindent
In summary, the relevance of
accurate theoretical predictions for the precise
analysis of slepton properties at a future linear collider has been outlined.
The full next-to-leading order corrections to the production of selectrons
and smuons were presented and shown to be sizeable, including potentially large
non-decoupling effects from superpartners in the loops.

\end{document}